\documentclass[aps,pre,preprint,showpacs,eqsecnum]{revtex4}
\usepackage{amsmath,bm,graphicx}

\begin{document}

\title{
Effective dynamics and steady state of an Ising model submitted to tapping
processes}

\author{A.\ Prados}
\email[]{prados@us.es}
\author{J.\ Javier Brey}
\email[]{brey@us.es}

\affiliation{
F\'{\i}sica Te\'orica, Universidad de Sevilla, Apartado de Correos 1065, E-41080
Sevilla, Spain}

\date{\today}

\begin{abstract}

A one-dimensional Ising model with nearest neighbour interactions is
applied to study compaction processes in granular media. An equivalent
particle-hole picture is introduced, with the holes being associated to
the domain walls of the Ising model. Trying to mimic the experiments, a
series of taps separated by large enough waiting times, for which the
system freely relaxes, is considered. The free relaxation of the system
corresponds to a $T=0$ dynamics which can be analytically solved. There is
an extensive number of metastable states, characterized by all the holes
being isolated. In the limit of weak tapping, an effective dynamics
connecting the metastable states is obtained. The steady state of this
dynamics is analyzed, and the probability distribution function is shown
to have the canonical form. Then, the stationary state is described by
Edwards thermodynamic granular theory. Spatial correlation functions in
the steady state are also studied.
\end{abstract}

\pacs{81.05.Rm, 05.50+q, 45.70.Cc}

\maketitle

\section{Introduction}
\label{s0}

Granular systems have attracted the attention of physicists in recent
years. A review of some features of granular matter can be found in
Refs.~\onlinecite{JNyB96,Ka98}. One of the most outstanding problems is
the phenomenon of compaction, i.\ e.\/, the increase of the density of a
loosely packed granular system when submitted to vibration or, more
generally, to some kind of external excitation. Compaction has been
extensively analyzed in a series of experiments by the Chicago group
\cite{KFLJyN95,NKPJyN97,NKBJyN98}. Starting from a low-density
configuration, near the random loose packed state, a system of
monodisperse glass beads was vertically tapped. Between taps, a long
enough time was considered, so that the system reached a mechanically
stable (metastable) configuration before the next tap started. The density
was measured in the metastable states, and its evolution as a function of
the number of taps was studied. The parameter controlling the dynamics of
the system is the dimensionless vibration intensity $\Gamma=a/g$, where
$a$ is the peak acceleration in the tap, and $g$ is the gravity. The
density was observed to increase very slowly towards a steady value
following an inverse logarithmic law \cite{KFLJyN95,NKPJyN97},  the steady
density being a monotonic decreasing function of the vibration intensity
\cite{NKBJyN98,Ja98}. Several models, with different underlying physical
mechanisms, have been proposed to understand this behaviour
\cite{CLHyN97,NCyH97,BKNJyN98,BPyS99,ByM01,SLyM02}, but a complete and
detailed physical theory is still lacking.

Simple Ising systems are often used as a first approximation to many
different phenomena in statistical physics. In a recent work
\cite{LyD01,DyL01}, the one-dimensional Ising model with nearest neighbour
interactions has been applied to analyze the problem of compaction in
dense granular media. This system can be regarded as one of the simplest
cases of spin models on random graphs, which have been very recently used
to investigate different aspects of granular matter
\cite{ByS01,ByM02,ByM01}. The ``tapping'' process is simulated in the
following way, in order to mimic what is done in the experiments
\cite{KFLJyN95,NKBJyN98,NKPJyN97}. First, the system freely relaxes until
it gets stuck in a metastable configuration ($n=0$). This is done by
considering a modified Glauber dynamics at $T=0$, in which only those
transitions lowering the energy of the system are permitted. Thus, all the
configurations with no spins antiparallel to both of their nearest
neighbours are metastable, i.\ e.\/, they are absorbent states \cite{vK92}
of this $T=0$ dynamics. Starting from the metastable state $n=0$, the
system is ``vibrated'' by allowing each spin to flip with probability $p$,
independently of the state of its neighbours. Afterwards, another free
relaxation at $T=0$ is considered, and the system reaches a new metastable
configuration $n=1$. By repeating this process, a chain $n=0,1,2,\ldots$
of metastable configurations is generated. It is found that the energy of
the system is a monotonic decreasing function of the number of taps $n$.
This Ising model can be mapped  on a particle-hole model, in which a
particles is associated to site $i$ if spins $i$ and $i+1$ are parallel,
and a hole is spins $i$ and $i+1$ are antiparallel. In this way, holes
correspond to the domain walls between arrays of parallel spins.
Interestingly, the dynamics at $T=0$ can be analytically solved in the
particle-hole picture \cite{PyB01a}. As a decrease in the energy
corresponds to a decrease in the number of domain walls, the density of
holes is a monotonic decreasing function of the number of taps $n$  in the
tapping process, i.\ e.\/, compaction takes place. In the reminder of the
paper, we will refer to this system as the TIM (tapped Ising model).

An analogous description of tapping processes was previously introduced
for the one-dimensional one-spin facilitated Ising model (1SFM)
\cite{BPyS99,BPyS00}, originally proposed by Fredrickson and Andersen in
the context of structural relaxation in glassy systems \cite{FyA84}. In
this model there are no interactions, but only an applied external field
$h$, and a spin can only flip if at least one of its nearest neighbours is
in the excited state. This system is also equivalent to a particle-hole
model, in which particles are associated to the spins aligned with the
field, while holes correspond to the spins in the excited state. Then,
facilitated spin flips are equivalent to adsorption and desorption of
particles on the one-dimensional lattice. These processes can only occur
provided that there is a hole on at least one of the nearest neighbour
sites of the flipping spin. The tapping process is modelled in the
following way \cite{BPyS99}. First, the system relaxes at $T=0$, where
only adsorption processes are allowed, until the system reaches a
metastable configuration $n=0$. Due to the facilitation rule, the
metastable states are characterized by all the holes being isolated, i.\
e.\/, surrounded by two particles. Then, starting from the metastable
state $n=0$, vibration is introduced by letting the system evolve, with
both facilitated adsorption and desorption, for a given time $t_0$.
Afterwards, the system relaxes again at $T=0$, which leads the system to a
new metastable state $n=1$. Iteration of this process gives a set
$n=0,1,2, \cdots$ of metastable configurations, with the density of
particles being an increasing function of the number of taps $n$
\cite{BPyS99,BPyS00}.

One of the most interesting physical questions in the problem of
compaction is the description of the steady state reached by the system in
the limit of an infinite number of taps. Note that thermal energy is
irrelevant for granular systems. The  important energy scale for a grain
of mass $m$ and diameter $d$ is $mgd$, where $g$ is the gravity. In a
typical granular system, $mgd/k_B T\simeq 10^{12}$ at room temperature.
Therefore, while molecular systems explore phase space due to thermal
fluctuations, in a powder thermal energy is negligible. Unless the system
is externally perturbed, each metastable configuration would last
indefinitely. Thus, thermodynamics is not directly applicable to powders.
Nevertheless, some years ago, Edwards and coworkers \cite{EyO89,MyE89}
made the hypothesis that the steady state of an externally perturbed
granular system can be described by an extension of the usual statistical
mechanics concepts to granular media. The central point is the ergodic
hypothesis for externally perturbed powders: in the steady state, all the
metastable configurations of a granular assembly occupying the same volume
are equiprobable. Besides, its most stable configuration corresponds to
the minimum volume. Therefore, the volume of a granular system is the
analog to the energy for a molecular system.  The entropy is defined as
the logarithm of the number of metastable configurations, which is
expected to be an extensive quantity. Then, it is possible to define a new
parameter, the compactivity $X=\partial V/\partial S$, playing the role of
the temperature in a molecular system, with  the limit $X=0$ giving the
most compact state.

In the  last years, a lot of effort have been carried out in order
to understand if the above ``equilibrium statistical mechanics''
or ``thermodynamic'' approach describes accurately the steady
state of an externally perturbed granular system. Most of this
effort has been focused on the analysis of simple models
\cite{LyD01,DyL01,BPyS00,PByS00,BFyS02,BKLyS00,BKLyS01,Le02},
although there has been also some attempts to test Edwards' theory
in experiments with real granular systems \cite{NKBJyN98}. Very
recently,  the theory has been checked in a numerical experiment
with a realistic granular matter model, specially conceived to be
reproducible in the laboratory \cite{MyK02}. In the context of
simple models, there is some numerical evidence of the validity of
the thermodynamic description in the limit of gently tapped
systems \cite{BKLyS00,BKLyS01,LyD01,BFyS02,Le02}, although for
stronger tapping the situation is not clear. In fact, numerical
results \cite{Le02,BFyS02} show that in the limit of strong
tapping the Edwards measure does not provide an accurate
description of the stationary state, at least in some spin models.
This has suggested an extension of the Edwards approach by
introducing a ``restricted'' measure \cite{Le02,BFyS02}. On the
other hand, analytical results are scarce, even for the simplest
models. To the best of our knowledge, the only system in which
Edwards theory has been analytically derived for tapping dynamics
is the 1SFM described above, in the limit of weak tapping
\cite{BPyS00}. An effective dynamics for the tapping process,
connecting metastable configurations, was obtained and the steady
probability distribution was shown to have the canonical form.
This leads to a relationship between Edwards' compactivity and the
dimensionless vibration intensity. Let us also mention that
Crisanti {\em et al.} \cite{CRRyS00} have studied one-dimensional
kinetically restricted models to address the validity of the
Stillinger-Weber construction, an approach that is related to the
Edwards measure.

Due to the lack of analytical results, it seems interesting to investigate
the possibility of deriving a thermodynamical description in the steady
state of other simple models. This is an important task from a theoretical
point of view. Firstly, it is a relevant question if the thermodynamic
picture is valid or not for models reproducing the experimentally observed
behaviour in granular systems. Secondly, if the answer is positive, it
might be possible to derive relationships between the parameters
controlling the evolution of the system, for instance the tapping
intensity $\Gamma$ in the compaction experiment, and the compactivity $X$,
which characterizes the stationary state. In this paper, we will center on
the analysis of the effective dynamics and the steady state of the TIM,
also in the limit of a gently tapped system. We will obtain the effective
dynamics between metastable states as an analytical approximation to the
original tapping dynamics. This will allow us to derive, also
analytically, the steady state probability distribution.

The paper is organized as follows. In Sec.\ \ref{s1} the model is
introduced, while the analytical solution of the modified Glauber dynamics
at $T=0$ is presented in Sec.\ \ref{s2}. Section \ref{s3} is devoted to
the derivation of the effective dynamics in the limit of weak tapping. The
properties of the steady state reached in the limit of an infinite number
of taps are discussed in Sec.\ \ref{s4}. It is shown that the steady
probability can be written in the canonical form, with the role of the
energy played by the volume and the temperature being substituted by a new
parameter, the compactivity, which is related to the tapping intensity.
Then, the steady state follows the statistical mechanics theory of Edwards
and coworkers \cite{EyO89,MyE89}. Finally, Sec.\ \ref{s5} contains a
summary of the work and some final remarks.

\section{An Ising model at $\bm{T=0}$ (free relaxation)}
\label{s1}

The hamiltonian of the one-dimensional Ising model is given by
\begin{equation}\label{1.1}
        {\cal H}=-J \sum_{i=1}^N \sigma_i \sigma_{i+1} \, ,
\end{equation}
where $J>0$ is the ferromagnetic coupling constant, $N$ is the number of
spins on the lattice, and $\sigma_i=\pm 1$ is the spin variable at site
$i$. We will consider periodic boundary conditions, so that formally
$\sigma_{N+1}=\sigma_1$. The time evolution of the system is governed by
single-spin-flip dynamics \cite{Gl63}. The probability $p(\bm{\sigma},t)$
for finding the system in configuration $\bm{\sigma}\equiv\{\sigma_i \}$
at time $t$ obeys a master equation of the form
\begin{equation}\label{1.2}
  \partial_t p(\bm{\sigma,t}) =\sum_{i=1}^N \left[
  w (\bm{\sigma}|R_i\bm{\sigma}) p(R_i\bm{\sigma},t)-
  w (R_i\bm{\sigma}|\bm{\sigma}) p(\bm{\sigma},t) \right] \, .
\end{equation}
Here $R_i\bm{\sigma}$ is the configuration obtained from $\bm{\sigma}$ by
just changing the state of spin $i$, and
$w(\bm{\sigma}|\bm{\sigma^\prime})$ stands for the transition rate from
configuration $\bm{\sigma^\prime}$ to $\bm{\sigma}$. Following Lefevre and
Dean \cite{LyD01,DyL01}, we introduce a $T=0$ dynamics such that only
those spin flips decreasing the energy of the system are allowed. Namely,
the transition rates are
\begin{equation}\label{1.3}
  w(R_i\bm{\sigma}|\bm{\sigma})=\frac{\alpha}{4} \left(
  1-\sigma_{i-1}\sigma_i \right) \left(1-\sigma_i\sigma_{i+1}\right)
  \, .
\end{equation}
It is easily verified that the above expression vanishes unless spin $i$ is
antiparallel to both of its nearest neighbours. The constant $\alpha$ defines
the basic time scale of the system. Interestingly, the most general transition
rates bringing the Ising model to equilibrium at temperature $T$ are \cite{Gl63}
\begin{eqnarray}
   w(R_i\bm{\sigma}|\bm{\sigma})& =&\frac{\alpha}{4} \Bigg[
   1+\delta \sigma_{i-1} \sigma_{i+1}  \nonumber \\
    &  & - \frac{1+\delta}{2}
   \sigma_i \left(\sigma_{i-1}+\sigma_{i+1}\right) \tanh \left(
   \frac{2J}{k_B T} \right) \Bigg] \, , \nonumber \\
   \label{1.4} & &
\end{eqnarray}
where $\delta$ is an arbitrary constant. The usual Glauber dynamics
corresponds to the choice $\delta=0$, whereas Eq.~(\ref{1.3}) is the zero
temperature limit of the case $\delta=1$. The dynamics defined by the
transition rates (\ref{1.3}) cannot be solved in the standard way
\cite{Gl63}, since the hierarchy of equations for the moments $C_n=\langle
\sigma_i\sigma_{i+1}\rangle$ is not closed.

Let us go to an equivalent description of the Ising model in terms of particles
and holes, by introducing a new set of variables
\begin{equation}\label{1.5}
  m_i=\frac{1-\sigma_i\sigma_{i+1}}{2} \, .
\end{equation}
If spins at sites $i$ and $i+1$ are antiparallel, it is $m_i=1$ and we
will refer to the site $i$ as empty or, equivalently, as being occupied by
a hole. On the other hand, if spins at site $i$ and $i+1$ are parallel it
is $m_i=0$, and site $i$ is occupied by a particle. Therefore, holes are
associated to the boundaries between domains of parallel spins, i.\ e.\/,
to the  so-called domain walls of the system. Due to the periodic boundary
conditions, the number of holes must be even in any configuration.

In terms of the $m_i$ variables, the hamiltonian (\ref{1.1}) reads
\begin{equation}\label{1.6}
  {\cal H}(\bm{m})=-J\sum_{i=1}^N \left( 1-2 m_i \right)=-JN+2J\sum_{i=1}^N m_i
  \, .
\end{equation}
A dimensionless energy per spin $\varepsilon$ can be defined as
\begin{equation}\label{1.7}
  \varepsilon=\frac{\langle {\cal H}\rangle_t}{JN}=-1+\frac{2}{N}\sum_{i=1}^N
  \langle m_i\rangle_t \, ,
\end{equation}
where we use the notation
\begin{equation}\label{1.8}
  \langle A(\bm{m}) \rangle_t =\sum_{\bm{m}} A(\bm{m}) p(\bm{m},t)
  \, ,
\end{equation}
for an arbitrary function $A(\bm{m})$ of the site variables $m_i$.

In the hole-particle description of the dynamics, the flip of spin $i$ modifies
both the values of $m_{i-1}$ and $m_i$. Then, from Eq.~(\ref{1.3}) we get for
the transition rates in the $\bm{m}$ variables
\begin{equation}\label{1.9}
  w(R_{i-1}R_i \bm{m}|\bm{m})=\alpha m_{i-1} m_i \, ,
\end{equation}
where $R_i$ is now the operator which transforms $m_i$ into $1-m_i$. The
master equation for the probability $p(\bm{m},t)$ is
\begin{eqnarray}
  \partial_t p(\bm{m},t) & = & \sum_i \left[ w(\bm{m}|R_{i-1}R_i\bm{m})
  p(R_{i-1}R_i\bm{m},t) \right. \nonumber \\
  \label{1.10} & & \left. -w(R_{i-1}R_i\bm{m}|\bm{m}) p(\bm{m},t)
  \right] \, .
\end{eqnarray}
In the dynamics defined by Eqs.~(\ref{1.9}) and (\ref{1.10}), the only
possible transitions are the simultaneous adsorption of two particles on
any two neighbouring empty sites. After a long enough time, the system
becomes trapped in a metastable state characterized by all the holes being
isolated, i.\ e.\/, all the empty sites surrounded by two particles. Of
course, the particular metastable state reached by the system will depend
on the initial conditions.

The present model displays some similarities as compared with the one
dimensional one-spin facilitated Ising model (1SFM) \cite{FyA84} at $T=0$. In
the latter, a particle can be adsorbed on an empty site as long as at least one
of its nearest neighbouring sites is empty \cite{BPyS99,BPyS00}.  Although the
dynamics of both models are not equivalent, the metastable states are the same,
being characterized by having all the empty sites isolated.

\section{Analytical solution of the dynamics at $\bm{T=0}$}
\label{s2}

Let us define the set of moments
\begin{equation}\label{2.1}
  D_r(t)=\langle m_k m_{k+1} \cdots m_{k+r}\rangle_t \, ,
\end{equation}
with $r\geq 0$. In the following, we will restrict ourselves to homogeneous
states, so that $D_r(t)$ does not depend on the position $k$ of the first site
considered. The lowest moment
\begin{equation}\label{2.2}
  D_0(t)=\langle m_k\rangle_t \,
\end{equation}
is the density of holes. This quantity can be related to the energy per particle
by means of Eq.~(\ref{1.7}),
\begin{equation}\label{2.3}
  \varepsilon=-1+2 D_0 \, .
\end{equation}
A hierarchy of equations for the moments $D_r(t)$ is easily obtained from the
master equation (\ref{1.10}),
\begin{equation}\label{2.4}
  \partial_t D_r(t)=-\alpha r D_r(t) -2\alpha D_{r+1}(t) \, ,
\end{equation}
valid for all $r\geq 0$. In order to solve the above hierarchy we
introduce the generating function
\begin{equation}\label{2.5}
  G(x,t)=\sum_{r=0}^\infty \frac{x^r}{r!} D_r(t) \, ,
\end{equation}
from which all the moments $D_r(t)$ can be obtained through
\begin{equation}\label{2.6}
  D_r(t)=\left[ \frac{\partial^r G(x,t)}{\partial x^r} \right]_{x=0} \, .
\end{equation}
From Eq.~(\ref{2.4}), it follows that the function $G(x,t)$ obeys the
first-order partial differential equation
\begin{equation}\label{2.7}
  \partial_t G(x,t)+\alpha (x+2) \partial_x G(x,t)=0 \, .
\end{equation}
By using standard techniques, the general solution of the above equation
is found to be
\begin{equation}\label{2.8}
  G(x,t)=G_0\left[ (x+2)e^{-\alpha t} - 2\right] \, ,
\end{equation}
where the function $G_0(y)$ is the initial condition, i.\ e.\/,
\begin{equation}\label{2.9}
  G_0(y)\equiv G(y,0)= \sum_{r=0}^\infty \frac{y^r}{r!} D_r(0) \, .
\end{equation}
In the long time limit it is
\begin{equation}\label{2.10}
  G(x,\infty)=G_0(-2) \, ,
\end{equation}
so that
\begin{equation}\label{2.11}
  \lim_{t\rightarrow\infty} D_0(t)=G_0(-2)
\end{equation}
and
\begin{equation}\label{2.12}
  \lim_{t\rightarrow\infty} D_r(t)=0
\end{equation}
for $r\geq 1$. The last result shows that all the holes become isolated in
the long time limit, and the probability of finding $r+1$ consecutive
holes, which equals $D_r$, vanishes for $r\geq 1$. The asymptotic density
of holes depends on the initial state, as indicated by Eq.~(\ref{2.11}).
In fact, it is trivially seen that the hierarchy of equations (\ref{2.4})
admits as a solution any constant value of $D_0$ as long as $D_r=0$ for
$r\geq 1$.

Now, let us specify the initial condition. We will consider that the
system is in equilibrium at temperature $T$ at $t=0$. The equilibrium
distribution is given by the canonical distribution corresponding to the
hamiltonian (\ref{1.1}),
\begin{equation}\label{2.13}
  p_T^{\text{eq}}(\bm{\sigma})=\frac{e^{-\beta {\cal H}(\bm{\sigma})}}
                       {\sum_{\bm{\sigma}} e^{-\beta {\cal H}(\bm{\sigma})}}
\end{equation}
or, equivalently,
\begin{equation}\label{2.14}
   p_T^{\text{eq}}(\bm{m})=\frac{e^{-\beta {\cal H}(\bm{m})}}
                       {\sum_{\bm{m}} e^{-\beta {\cal H}(\bm{m})}} \, ,
\end{equation}
where $\beta=(k_B T)^{-1}$. Using the hamiltonian (\ref{1.6}), it is easy to
show that
\begin{subequations}
\begin{equation}\label{2.15a}
  p_T^{\text{eq}}(\bm{m})=\prod_{i=1}^N p_T^{\text{eq}} (m_i) \, ,
\end{equation}
\begin{equation}\label{2.15b}
  p_T^{\text{eq}}(m_i)=\frac{e^{-2 \beta J m_i}}{1+e^{-2\beta J}} \, .
\end{equation}
\end{subequations}
Therefore,
\begin{subequations}
\begin{equation}\label{2.16a}
  D_{0,T}^{\text{eq}}=\langle m_k ; T\rangle_{\text{eq}}=
  \frac{e^{-2\beta J}}{1+e^{-2\beta J}} = a \, ,
\end{equation}
\begin{equation}\label{2.16b}
  D_{r,T}^{\text{eq}}=\langle m_k m_{k+1} \cdots m_{k+r} ;T\rangle_{\text{eq}}=
  \langle m_k ; T\rangle_{\text{eq}}^{r+1}=a^{r+1} \, ,
\end{equation}
\end{subequations}
where $0\leq a\leq 1$. The value $a=0$ corresponds to $\beta\rightarrow\infty$
($T\rightarrow 0^+$), and $a=1$ to $\beta\rightarrow -\infty$ ($T\rightarrow
0^-$). In the limit $\beta\rightarrow 0$ ($T\rightarrow\infty$), it is $a=1/2$,
which corresponds to the most disordered state.

Therefore, the initial condition corresponding to an equilibrium state is
given by
\begin{equation}\label{2.17}
  D_r(0)=a^{r+1} \, ,
\end{equation}
which leads to
\begin{equation}\label{2.18}
  G_0(x)\equiv G(x,0)=\sum_{r=0}^\infty \frac{x^r}{r!} a^{r+1}=a e^{a x} \, .
\end{equation}
With this choice, the solution given by Eq.~(\ref{2.8}) takes the form
\begin{equation}\label{2.19}
  G(x,t)=a \exp \left\{ a \left[ (x+2)e^{-\alpha t} -2 \right] \right\}
\end{equation}
and, consequently, by using Eq.~(\ref{2.6}),
\begin{equation}\label{2.20}
  D_r(t)=   a^{r+1} \exp \left[ -\alpha r t+ 2 a \left( e^{-\alpha t}-1
  \right)\right]
  \, .
\end{equation}
As pointed out above, all the moments $D_r$ with $r\geq 1$ vanish in the long
time limit, while the asymptotic value of the density of holes reads
\begin{equation}\label{2.21}
  \lim_{t\rightarrow\infty}D_0(t)=a e^{-2 a} \, ,
\end{equation}
which depends on the initial density of holes $a$, being always smaller than it,
since only adsorption processes are allowed in the $T=0$ dynamics. The
dimensionless energy per particle in the metastable final configuration
$\varepsilon_\infty$ follows directly from Eqs.~(\ref{2.3}) and (\ref{2.21}),
\begin{equation}\label{2.22}
  \varepsilon_\infty=-1+2 a e^{-2 a} \, .
\end{equation}
This expression agrees with the result obtained by Lefevre and Dean
\cite{LyD01,DyL01}. The energy $\varepsilon_\infty$ is maximum for
$a=1/2$, i.\ e.\/, when the system starts from a fully random
configuration. Of course, this is equivalent to say that $D_0(\infty)$ has
a maximum for $a=1/2$. The existence of this maximum is in contrast with
the result for the 1SFM, where the  asymptotic density of holes is a
monotonic function of the initial density \cite{FyR96,BPyS99}.

Therefore, at $T=0$ the following picture emerges. Starting from any
configuration, the system evolves until all the holes become isolated, i.\
e.\/, it gets stuck in a metastable state characterized by all the moments
$D_r$ vanishing for $r\geq 1$. Going back to the spin description, the
metastable states are those such there is no spin antiparallel to both of
its nearest neighbours. In other words, all the domains of parallel spins
have, at least, a length of two sites. The density of holes in the
metastable state, or the density of domain walls in the spin image,
depends on the initial configuration, being given by Eq.~(\ref{2.21}).

\section{Effective dynamics for tapping processes}
\label{s3}

Let us consider the model introduced in section  \ref{s1} to get a
theoretical approach to the compaction processes in vibrated granular
systems. The model is tapped in the following way \cite{LyD01,DyL01},
trying to mimic what is done in the experiments with real granular
materials \cite{KFLJyN95,NKPJyN97,NKBJyN98}. Firstly, the system freely
relaxes as described in the previous section, until getting trapped in a
metastable configuration ($n=0$), characterized by the absence of spins
being antiparallel to {\it both} of their nearest neighbours. These
configurations are absorbent states \cite{vK92} for the dynamics at $T=0$.
Secondly, starting from the metastable configuration $n=0$, the system is
``vibrated''. Each spin can flip with a probability $p$, independently of
the state of its neighbours. Afterwards, another free relaxation at $T=0$
follows, until the system becomes again stuck in a new metastable
configuration, $n=1$. By repeating this process, a chain of metastable
configurations $n=0,1,2,\ldots$ is generated. It is important to note that
we can restrict ourselves to values of $p$ in the interval $0\leq p\leq
1/2$, since the same evolution of the energy is obtained for both $p$ and
$1-p$. This is because a probability $1-p$ is equivalent to a simultaneous
flip of all the spins (which does not change the energy), followed by a
flip of each spin with probability $p$. Making use of the notation
introduced in the introduction, we will refer to this tapped
one-dimensional Ising model as the TIM.

If $p\ll 1$, the evolution of the system is very slow, since it is very
improbable that a given spin flips. The dynamics will be dominated by
those transitions in which only a few spins change their state during a
vibration process. Therefore, for $p\ll 1$ an expansion in powers of $p$
may be useful, since we hope that  retaining  the lowest orders would
provide a good approximation.

In the previous section we have shown that, in the description of
particles and holes, the metastable states are characterized by having all
the holes isolated, i.\ e.\/, there are no domains of parallel spins with
length $l<2$. We are going to consider the evolution of the system in a
single tap, defined as the sequence of a vibration process followed by a
free relaxation to a metastable configuration, to the lowest order in $p$.
In this limit, only one flip during the vibration takes place. The
analysis to be presented depends on the length $l$ of the domain of
parallel spins containing the flipping spin. Our goal is to obtain
expressions for the transition probabilities
$W_{\text{ef}}(\bm{m^\prime}|\bm{m})$ from the initial metastable
configuration $\bm{m}$ to the final one $\bm{m^\prime}$ in a single tap.

\begin{figure}
\includegraphics[scale=0.3]{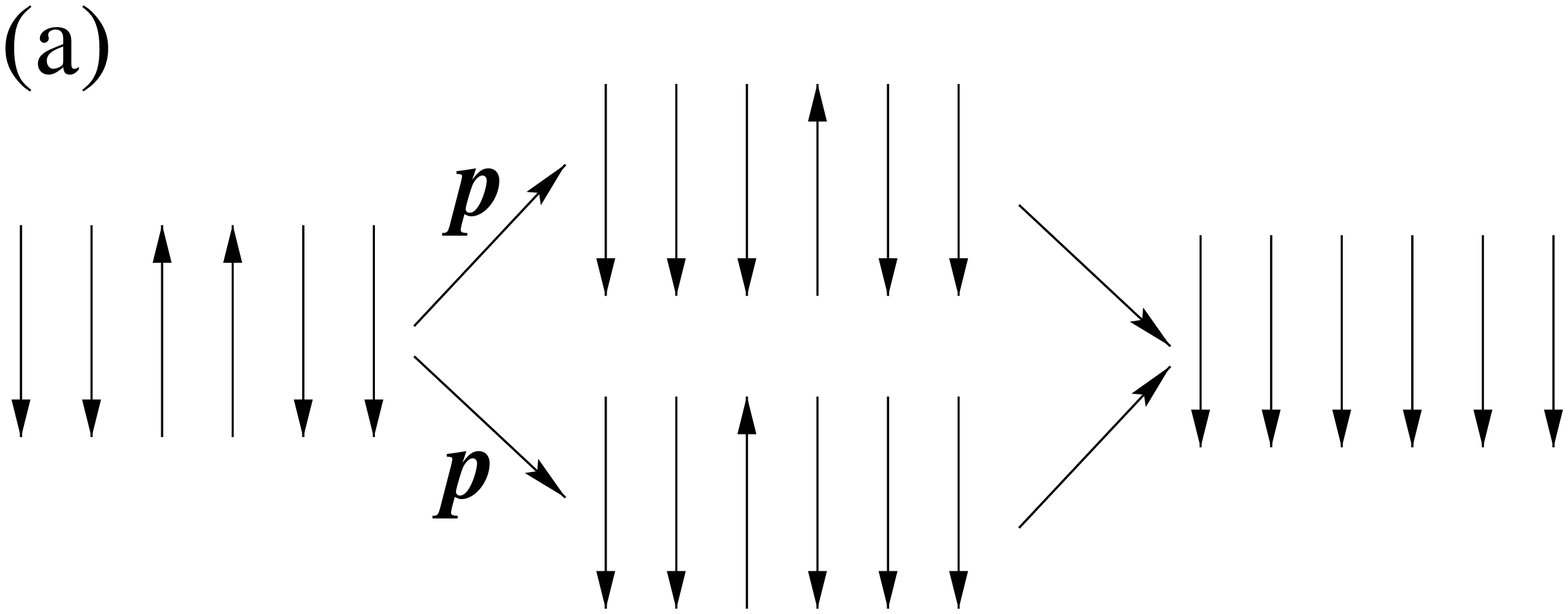}\\
\includegraphics[scale=0.3]{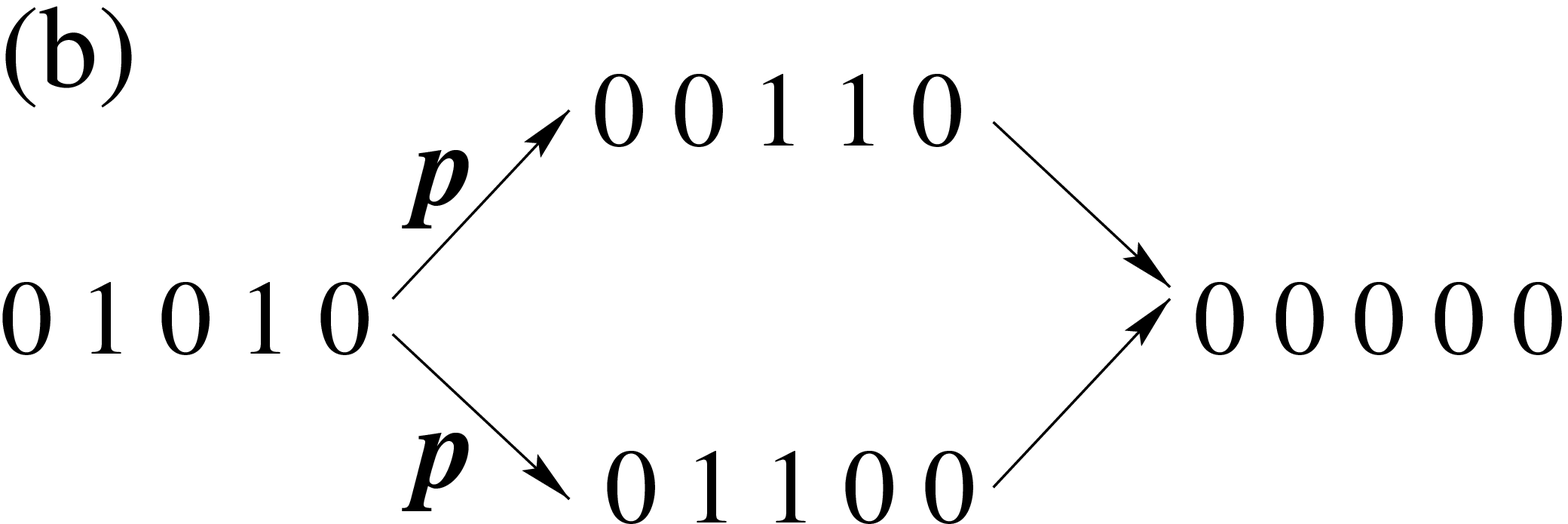}
\caption{\label{fig1}
(a) Destruction of a domain of length $l=2$ in the tapping process. In the
vibration, one of the spins of the domain flips with probability $p$, and the
remaining unstable domain of length $l=1$ disappears in the free relaxation at
$T=0$. (b) The same trajectories, in the particle and hole picture.}
\end{figure}

Let us assume first that the flipping spin belongs to a domain of
initial length $l=2$, i.\ e.\/, the transition is like
\begin{equation}\label{3.2}
   \ldots\downarrow\downarrow\underline{\uparrow}\uparrow\downarrow\downarrow
   \ldots \longrightarrow
   \ldots\downarrow\downarrow\underline{\downarrow}\uparrow\downarrow\downarrow
   \ldots \, ,
\end{equation}
where the flipping spin has been underlined. The probability that this
process occurs in the cluster above is $p(1-p)^5$. Afterwards, in the free
relaxation at $T=0$ the up-spin has to flip necessarily, reaching a new
metastable state in which the domain with $l=2$ has disappeared.
Therefore, to first order in $p$
\begin{equation}\label{3.3.}
  W_{\text{ef}}(\downarrow\downarrow\downarrow\downarrow\downarrow\downarrow|
        \downarrow\downarrow\uparrow\uparrow\downarrow\downarrow)
        = 2p \, ,
\end{equation}
where the factor of $2$ follows because of the other path connecting the
same initial and final states, and corresponding to the flip of the spin
on the right of the domain with $l=2$ during the vibration. These
trajectories are shown in FIG.~\ref{fig1}, both in the spin and in the
particle-hole pictures. Introducing an usual notation, in the transition
rates $W_{\text{ef}}(\bm{m^\prime}|\bm{m})$ we have only indicated the
sites involved in the given rearrangement. In the particle and hole
picture, it is
\begin{equation}\label{3.4}
  W_{\text{ef}}(00000|01010)=2p \, ,
\end{equation}
the process involves the elimination of both holes. During the vibration,
one hole diffuses next to the other one, so that in the free relaxation
two particles are simultaneously adsorbed on them, as it is shown in
FIG.~\ref{fig1}.

When the flipping spin belongs to a domain of length $l>2$, we have to
analyze two cases separately, by distinguishing whether the flipping spin
during the vibration is located at the domain wall or not. In the former
case, for instance
\begin{equation}\label{3.6}
   \ldots\downarrow\downarrow\overbrace{\underline{\uparrow}\uparrow\uparrow\ldots}^l
   \longrightarrow
   \ldots\downarrow\downarrow\underline{\downarrow}\uparrow\uparrow\ldots \,
   ,
\end{equation}
no transition occurs in the following free relaxation, since the final
state in the vibration process is metastable. Therefore, the effective
transition probability between the metastable states is
\begin{equation}\label{3.7}
  W_{\text{ef}}(\downarrow\downarrow\downarrow\uparrow\uparrow|
        \downarrow\downarrow\uparrow\uparrow\uparrow) = p \, .
\end{equation}
Similarly,
\begin{equation}\label{3.8}
  W_{\text{ef}}(\uparrow\uparrow\downarrow\downarrow\downarrow|
        \uparrow\uparrow\uparrow\downarrow\downarrow)=p \, .
\end{equation}
These transitions are  one-site diffusion processes of a hole in both
directions,
\begin{equation}\label{3.9}
  W_{\text{ef}}(0010|0100)=W_{\text{ef}}(0100|0010)=p \, ,
\end{equation}
which are present for all $l>2$.

\begin{figure}
\includegraphics[scale=0.3]{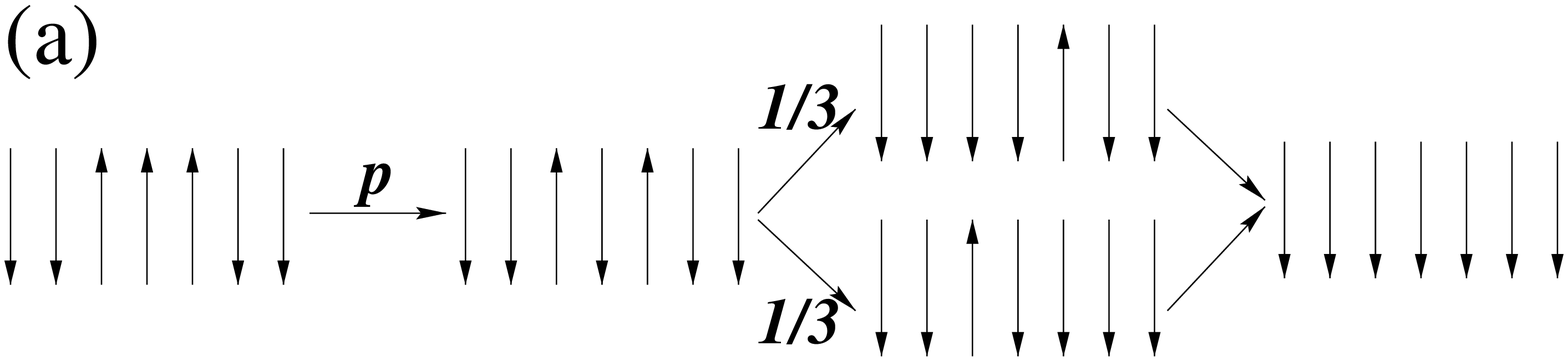}\\
\includegraphics[scale=0.3]{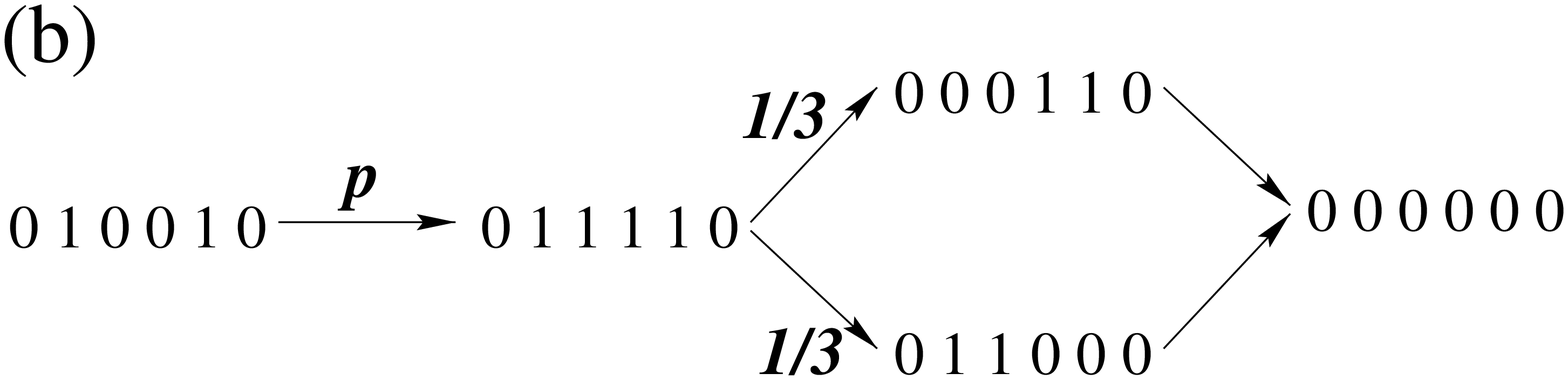}
\caption{\label{fig2} (a) Trajectories leading to the destruction of a
domain of length $l=3$ in a tapping process. In the vibration, the central
spin flips with probability $p$. Afterwards, the system freely relaxes at
$T=0$. Paths leading to a final state identical to the initial one are not
shown. (b) The same trajectories as in (a), in the particle-hole picture.
Note that the flip of one spin corresponds to the change of two
consecutive sites in the associated particle lattice.}
\end{figure}

When one of the internal spin flips, we have to analyze separately $l=3$
and $l\geq4$. First, let us consider $l=3$. The internal spin is a nearest
neighbour of both the spins at the domain walls, and the rearrangement
occurring in the vibration has the form
\begin{equation}\label{3.10}
  \ldots\downarrow\downarrow\overbrace{\uparrow\underline{\uparrow}\uparrow}
  \downarrow\downarrow\ldots
  \longrightarrow
  \ldots\downarrow\downarrow\overbrace{\uparrow\underline{\downarrow}\uparrow}
  \downarrow\downarrow\ldots \,\,\, .
\end{equation}
Afterwards, in the free relaxation, any of the three spins of the domain
can flip, with the same probability, i.\ e.\/, $1/3$. If it is the central
spin the one flipping, returning then to its original state, nothing has
occurred globally, and the group is again in the initial state. On the
other hand, if any of the external spins of the domain flips first, say
the one on the left, the reached state is not metastable yet. Then, the
free relaxation involves another flip, in which the upward spin has
necessarily to go down. Therefore, the effective transition probability
for the complete process is
\begin{equation}\label{3.12}
  W_{\text{ef}}(\downarrow\downarrow\downarrow\downarrow\downarrow\downarrow
  \downarrow|\downarrow\downarrow\uparrow\uparrow\uparrow\downarrow\downarrow)
  = \frac{2}{3} p \, .
\end{equation}
The trajectories leading to this rearrangement are shown in
FIG.~\ref{fig2}. The factor of $2/3$ appears because the final state is
the same, independently of which is the first external spin flipping in
the free relaxation, and each trajectory in FIG.~\ref{fig2} contributes
$p/3$. In the particle-hole picture, the process consists of the
occupation of two holes separated by two particles, with a probability
\begin{equation}\label{3.13}
  W_{\text{ef}}(000000|010010)=\frac{2}{3}p  \, .
\end{equation}

In a domain of length $l\geq 4$, there are $l-2$ internal spins. Suppose
that the one next to the left wall flips in the vibration,
\begin{equation}\label{3.14}
  \ldots\downarrow\downarrow\overbrace{\uparrow\underline{\uparrow}
  \uparrow\uparrow\ldots}^{l}
  \longrightarrow
  \ldots\downarrow\downarrow\overbrace{\uparrow\underline{\downarrow}
  \uparrow\uparrow\ldots}^{l}
\end{equation}
To the right of the flipping spin, a stable domain of length $l-2$
appears, and the spin to its left must move downwards in the free
relaxation, i.\ e.\/, the domain wall moves two sites to the right in the
whole process. The probability of the transition is
\begin{equation}\label{3.15}
  W_{\text{ef}}(\downarrow\downarrow\downarrow\downarrow\uparrow\uparrow|
  \downarrow\downarrow\uparrow\uparrow\uparrow\uparrow) = p \, ,
\end{equation}
or, in the particle-hole description,
\begin{equation}\label{3.16}
  W_{\text{ef}}(00010|01000)=p \, .
\end{equation}
The process corresponds to a two-sites diffusion of the hole to the right.
Obviously, it is also possible that a hole diffuses two sites to the left,
corresponding to the flip of the internal spin next to the right wall in
the vibration,
\begin{equation}\label{3.17}
  W_{\text{ef}}(01000|00010)=p \, .
\end{equation}
Finally, if the internal spin which flips in the vibration is not next to
any of  the domain walls, it has to return to its original state in the
free relaxation, and there is no global transition in the tap.

\begin{table}
\caption{
\label{tab1}
Probabilities of the first order transitions in a single tap, connecting
metastable states.}
\begin{ruledtabular}
\begin{tabular}{cccc}
% Lines of table here ending with \\
{Process} & {Initial state} & {Final state} & $W_{\text{ef}}$ \\

{One-site diffusion} & 0100 & 0010 & $p$ \\

& 0010 & 0100 & $p$ \\

{Two-sites diffusion} & 01000 & 00010 & $p$ \\

& 00010 & 01000 & $p$
\\
Destruction of a hole pair & 01010 & 00000 & $2p$ \\

& 010010 & 000000 & $\frac{2}{3}p$
\end{tabular}
\end{ruledtabular}
\end{table}

The obtained transition probabilities up to first order in $p$ are
summarized in Table \ref{tab1}. The particle-hole description is used,
since it is  more convenient for the analysis of the compaction process.
Therefore, to the lowest order, with only one flip during the vibration,
the density of particles cannot decrease and compaction takes place. When
higher orders are retained, processes leading to a decrease of the number
of particles show up, as it will be discussed below.

The description of the effective dynamics between metastable states is not
complete, even qualitatively, if it is restricted to the lowest order in
$p$. In particular, the existence of a steady state characterized by the
flipping probability in the vibration $p$ \cite{LyD01,DyL01} is lost. The
Markov process describing the dynamics between metastable states is not
irreducible, and the configuration with all the sites being occupied by
particles is an absorbent state \cite{vK92} of the dynamics. Therefore, in
order to have a more complete description, we are led to consider higher
orders in $p$, i.\ e.\/, processes involving more than one transition
during the vibration. This will be done in a physical way, similar to that
of Ref.~\onlinecite{BPyS00}. We are not going to consider those second
order processes (two flips in the vibration) whose effect can be obtained
by means of a combination of two processes of order $p$, but only those
processes for which the effective transition probability $W_{\text{ef}}$
vanishes to the lowest order. In particular, this is the case for all
those trajectories decreasing the density of particles, as already
mentioned. The inclusion of these processes modifies essentially the
physics of the tapping process, so that they must be taken into account in
our effective dynamics.

Then, we will consider that there are two flips take place in the
vibration. As in the free relaxation only  transitions decreasing the
energy are allowed, we have to analyze two cases: (i) flip of two nearest
neighbours spins, and (ii) flip of two spins separated by one site. If the
two flipping spins are separated by more than one site, the local free
relaxations associated to each of them are independent, and the result is
a product of two first-order transitions.

\begin{figure}
\includegraphics[scale=0.3]{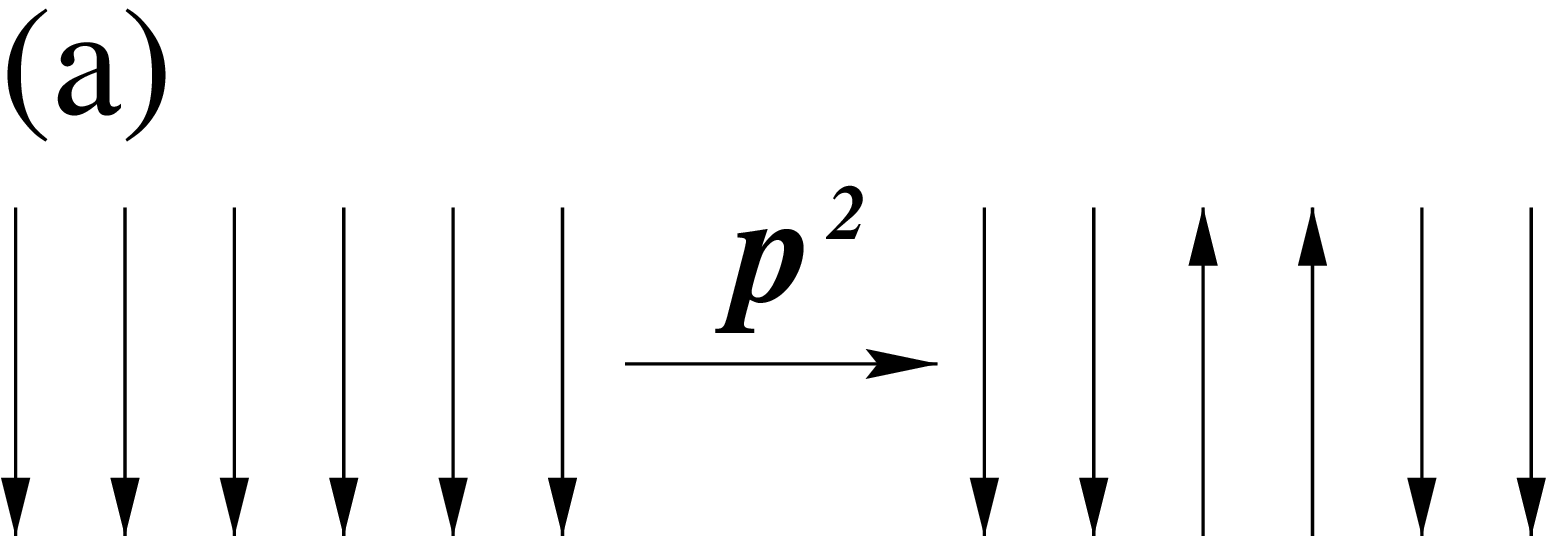}\\
\includegraphics[scale=0.3]{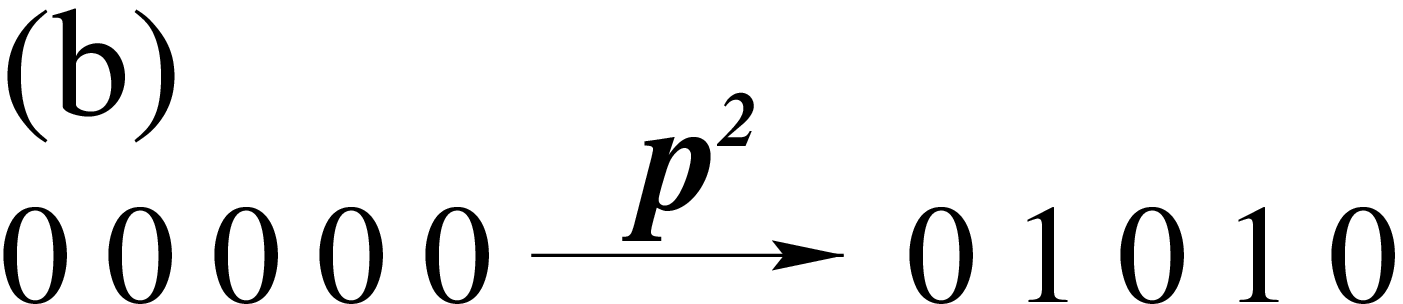}
\caption{\label{fig3}
(a) Creation of a domain of length $l=2$ in the tapping process. In the
vibration, two consecutive spins belonging to a domain of length $l\geq 6$ flip
with probability $p^2$. As a  consequence, a new stable domain of length $l=2$
shows up. (b) The preceding trajectory, in the particle-hole picture.}
\end{figure}

Suppose the transition during the vibration in the cluster depicted in
FIG.~\ref{fig3}, whose probability is $p^2(1-p)^4$. In the free relaxation
no transition can happen, since a metastable domain of length $l=2$ has
been created. Therefore, to the lowest order it is
\begin{equation}\label{3.19}
  W_{\text{ef}}(\downarrow\downarrow\uparrow\uparrow\downarrow\downarrow|
        \downarrow\downarrow\downarrow\downarrow\downarrow\downarrow)=p^2 \, ,
\end{equation}
or,  in the particle-hole picture,
\begin{equation}\label{3.20}
  W_{\text{ef}}(01010|00000)=p^2  \, .
\end{equation}
In order to derive this transition probability it has been assumed that
the domain which the flipping spins belong to initially is at least of
length $l=6$. If it has a smaller length, it is easily  shown that the
resulting rearrangement can be obtained as a combination of two
first-order processes.

\begin{figure}
\includegraphics[scale=0.3]{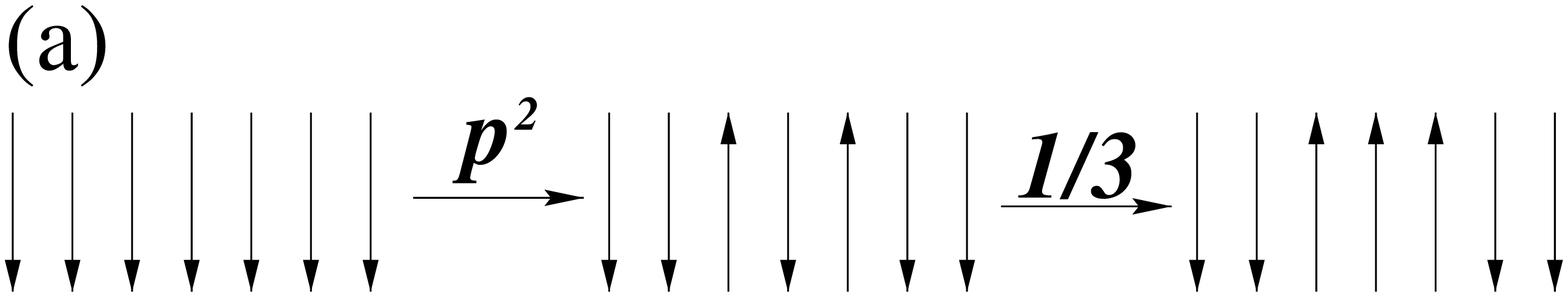}\\
\includegraphics[scale=0.3]{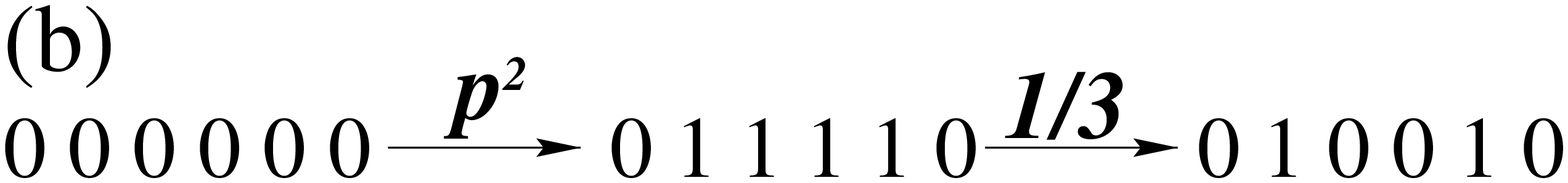}
\caption{\label{fig4}
(a) In the vibration, two spins of the domain of length $l\geq 7$ flip with
probability $p^2$, creating three consecutive unstable domains of length $l=1$.
If, in the subsequent free relaxation, it is the central spin the one which
flips, a new  stable domain of length $l=3$ appears. (b) The same trajectory,
shown for particles and holes.}
\end{figure}

The other kind of second order processes we have to include in our
approximation corresponds to a transition during the vibration of the form
\begin{equation}\label{3.21}
\ldots\downarrow\downarrow\underline{\downarrow}\downarrow\underline{\downarrow}
  \downarrow\downarrow\ldots
  \longrightarrow
   \ldots\downarrow\downarrow\underline{\uparrow}\downarrow\underline{\uparrow}
   \downarrow\downarrow\ldots
\end{equation}
whose probability,  to leading order, is $p^2$. In the free relaxation
process, first either one of the underlined spins which have just flipped
or the one between them flips, the three changes  having the same
probability. In the former case, the  group of spins returns to the
initial configuration, all of them  upwards. On the other hand, if the
central spin flips, a new domain of length $l=3$ appears, with probability
\begin{equation}\label{3.22}
W_{\text{ef}}(\downarrow\downarrow\uparrow\uparrow\uparrow\downarrow\downarrow|
  \downarrow\downarrow\downarrow\downarrow\downarrow\downarrow\downarrow)=
  \frac{1}{3}p^2 \, .
\end{equation}
The trajectory leading to this transition is shown in FIG.~\ref{fig4}. In
terms of  particles and holes
\begin{equation}\label{3.23}
  W_{\text{ef}}(010010|000000)=\frac{1}{3}p^2 \, .
\end{equation}

It is easy to convince oneself that there are no more second order
transitions increasing the number of holes, and that any other second
order process can be decomposed in a product of first-order transitions.
The new transitions appearing to second order are shown in Table
\ref{tab2}. Together with the first-order processes in Table \ref{tab1},
they define the approximate effective model for tapping processes we are
going to analyze in the following. It is important to note that the
introduction of the second order processes is fundamental from a physical
point of view. The processes described in  FIGS.~\ref{fig3} and
\ref{fig4}, with probabilities given by Eqs.~(\ref{3.20}) and
(\ref{3.23}), are the inverse of the processes in  FIGS.~\ref{fig1} and
\ref{fig2}, Eqs.~(\ref{3.4}) and (\ref{3.13}), respectively. Therefore,
the Markov process defined by the transition probabilities in Tables
\ref{tab1} and \ref{tab2} is irreducible \cite{vK92}, i.\ e.\/, all the
states are connected by a chain of transitions with non-zero probability.
This property will be fundamental in the analysis of the steady state,
presented in the next section.

\begin{table}
\caption{\label{tab2}
Probabilities of the second order processes between metastable states leading to
an increase of the number of holes.}
\begin{ruledtabular}
\begin{tabular}{cccc}
% Lines of table here ending with \\
Process & Initial state & Final state & $W_{\text{ef}}$ \\

Creation of a hole pair & 00000 & 01010 & $p^2$\\

& 000000 & 010010 & $\frac{1}{3} p^2$
\end{tabular}
\end{ruledtabular}
\end{table}

\section{Steady state solution}
\label{s4}

As pointed out above, the Markov process defined by the effective
transition probabilities connecting metastable states is irreducible
\cite{vK92} and, consequently, there is a unique steady state  for each
given value of $p$. This steady state will be reached by the system from
any initial configuration. Besides, it will be shown that the system
described by the effective master equation verifies detailed balance. By
using this property we will be able to obtain the steady distribution
analytically. With regards to the original model, the expression holds in
the limit of gently tapped systems, $p\ll 1$, for which the effective
transition probabilities of Tables \ref{tab1} and \ref{tab2} have been
obtained.

In order to calculate the steady distribution function, we will  bet a
priori on a stationary solution $P_s(\bm{m})$ of the master equation for
the tapping process verifying the detailed balance condition,
\begin{equation}\label{4.1}
  W_{\text{ef}}(\bm{m}|\bm{m^\prime}) P_s(\bm{m^\prime})=
  W_{\text{ef}}(\bm{m^\prime}|\bm{m}) P_s(\bm{m}) \, .
\end{equation}
Given the uniqueness of the steady state, if a solution is found in this
way, its own existence will be the proof of the detailed balance property
in the system. Detailed balance implies that all the configurations
$\bm{m}^{(k)}$ having the same number of holes $k$ will be equiprobable,
since they are connected through diffusion processes, which are isotropic.
Their probability will be denoted by $P_s(\bm{m}^{(k)})$. Moreover, for
the processes changing the density of the system it is
\begin{equation}\label{4.2}
\frac{P_s(\bm{m}^{\prime(k+2)})}
        {P_s(\bm{m}^{(k)})}=
\frac{W_{\text{ef}}(\bm{m}^{\prime(k+2)}|\bm{m}^{(k)})}
        {W_{\text{ef}}(\bm{m}^{(k)}|\bm{m}^{\prime(k+2)})}=\frac{p}{2} \,
        .
\end{equation}
This expression applies for both pairs of transitions with non-zero
probability, given by Eqs.~(\ref{3.20}) and (\ref{3.4}), and
Eqs.~(\ref{3.23}) and (\ref{3.13}), respectively (see also Tables
\ref{tab1} and \ref{tab2}). Consequently,
\begin{equation}\label{4.3}
  P_s(\bm{m}^{(k)})=\frac{1}{Z}\left( \frac{p}{2} \right)^{\frac{k}{2}} \, ,
\end{equation}
where $Z$ is a normalization constant, and we have taken into account that
the number of holes is always even. Defining a new variable $X$ by
\begin{equation}\label{4.4}
  e^{-1/X}=\sqrt{\frac{p}{2}} \, ,
\end{equation}
the steady probability distribution can be written in the ``canonical'' form
\begin{equation}\label{4.5}
  P_s(\bm{m}^{(k)})=\frac{1}{Z} e^{-k/X}      \, ,
\end{equation}
so that $X$ is identified as the compactivity of Edwards' statistical
mechanics theory of powders  \cite{EyO89,MyE89}. The number of holes $k$
plays the role of the volume or, more precisely, the excess volume from
the densest state. The normalization constant $Z$ is the analog to the
partition function. From it, all the steady properties of the system can
be obtained in the standard way.

The calculation of $Z$ is quite an easy task,
\begin{subequations}\label{4.6}
\begin{equation}\label{4.6a}
  Z=\sum_{k=0, \text{ $k$ even}}^{N/2}  Z_k  \, ,
\end{equation}
\begin{equation}\label{4.6b}
  Z_k=\Omega_{k}^{(N)} e^{-k/X} \, ,
\end{equation}
\end{subequations}
with $\Omega_{k}^{(N)}$ being the number of metastable states with $k$
holes for a lattice with $N$ sites. The maximum number of holes is $N/2$
(we are assuming that $N$ is even), and the number of holes $k$ must be
even in the TIM because of the periodic boundary conditions. A simple
combinatorial argument leads to
\begin{equation}\label{4.7}
  \Omega_k^{(N)}=\frac{N(N-k-1)!}{k! (N-2k)!} \, .
\end{equation}
In the large $N$ limit, the sum in Eq.~(\ref{4.6a}) can be evaluated by
the saddle point method, since $Z_k$ has a sharp maximum as a function of
$k$, with the result
\begin{equation}\label{4.8}
  \ln \zeta\equiv\frac{1}{N}\ln Z=\ln \frac{1+(1+4e^{-1/X})^{1/2}}{2} \, .
\end{equation}
The number of holes is the property analogous to the energy of a molecular
system, and the steady hole density reads
\begin{equation}\label{4.9}
  D_0^{s}=-\frac{\partial \ln\zeta}{\partial (1/X)}=
        \frac{(1+4e^{-1/X})^{1/2}-1}{2(1+4e^{-1/X})^{1/2}} \, .
\end{equation}
It must be stressed that $D_0^s=\bar{k}/N$, being $\bar{k}$ the value of the
number of holes $k$ for which $Z_k$ reaches its maximum. The steady probability
distribution is a very sharply peaked function around $\bar{k}$, which assures
the equivalence of the microcanonical and canonical ensembles for the
calculation of the mean values of the physical properties in the steady state.
The stationary density of holes $\rho^s=1-D_0^s$ is a monotonic decreasing
function of the compactivity $X$. As the compactivity,  given by
Eq.~(\ref{4.4}), increases with the vibration intensity $p$,  $\rho^s$ is also a
monotonic decreasing function of the vibration intensity, a behaviour analogous
to that of real granular systems \cite{NKBJyN98,Ja98}. In the limit $p\ll 1$,
Eq.~(\ref{4.9}) reduces to
\begin{equation}\label{4.10}
D_0^s \sim e^{-1/X}=\sqrt{p/2} \, .
\end{equation}

\begin{figure}
\includegraphics[scale=0.4]{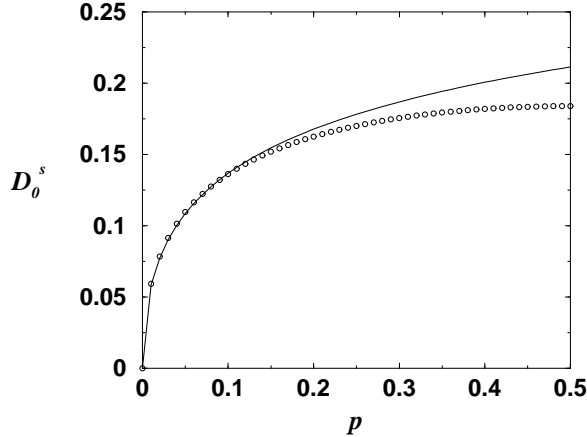}%
\caption{\label{fig5} Comparison of the numerical evaluation of the steady
density of holes, as given by Eq.~(\ref{4.11}), (circles) and the
analytical explicit expression obtained from the effective dynamics
approach, Eq.~(\ref{4.9}). The agreement is quite good for $p\lesssim
0.15$.}
\end{figure}

\begin{figure}
\includegraphics[scale=0.4]{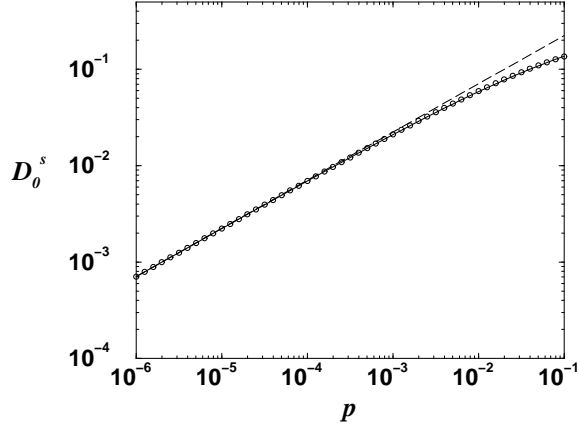}%
\caption{\label{fig6} Comparison of the numerical evaluation of the steady
density of holes, as given by the exact solution Eq.~(\ref{4.11})
(circles), the analytical expression obtained from the effective dynamics
approach, Eq.~(\ref{4.9}) (solid line), and their leading behaviour
(\ref{4.10}) in the limit $p\ll 1$ (dashed line), in the interval
$10^{-6}<p<10^{-1}$. It is observed that Eq.~(\ref{4.10}) only gives a
good approximation in the very weak tapping regime, $p\lesssim 10^{-3}$,
for which the steady density of particles $\rho^s=1-D_0^s\gtrsim 0.98$.}
\end{figure}

In Refs.~\onlinecite{LyD01,DyL01} it was found that the steady density of
holes for arbitrary $p$ is given by the solution of the equation
\begin{equation}\label{4.11}
   D_0^s=\left[2pq+D_0^s (1-4pq)\right] \exp\left[
  -\frac{2pq}{2pq+D_0^s(1-4pq)} \right] \, ,
\end{equation}
with $q=1-p$. Of course, as Eq.~(\ref{4.11}) gives the exact value of the
density of holes, it is symmetric against the change $p\rightarrow 1-p$.
In the limit $p\ll 1$, the above expression yields $D_0^s\sim \sqrt{p/2}$,
which agrees with the small $p$ limit of the expression obtained from the
effective dynamics, Eq.~(\ref{4.9}). In FIG.~\ref{fig5}, we compare the
density of holes as a function of $p$, obtained from Eqs.~(\ref{4.9}) and
(\ref{4.11}). It is observed that the agreement is quite good for
$p\lesssim 0.15$, i.\ e.\/, for a steady density of particles
$\rho^s=1-D_0^s\gtrsim 0.85$. On the other hand, the leading behaviour for
$p\ll 1$, Eq.~(\ref{4.10}), only holds for very small values of $p$,
$p\lesssim 10^{-3}$, as it is clearly shown in FIG.~\ref{fig6}. Therefore,
the accuracy of the results obtained from the effective dynamics picture
extends further than what might be expected from a second-order theory in
the flipping probability $p$. In fact, this is not so surprising, because
a similar behaviour was found in the one-dimensional facilitated Ising
model submitted to tapping processes \cite{BPyS00}. The wide range of
applicability presented by Eq.~(\ref{4.9}) can be understood on physical
grounds, by realizing that the effective dynamics approach is not a
standard second-order expansion in the flipping probability $p$. Among all
the second-order processes, only those introducing new physically relevant
transitions, i.\ e.\/, transitions that cannot be written as a combination
of first-order processes, are collected. This physically motivated
expansion allows an analytical approach to the rather difficult problem of
tapping.

The entropy is defined in the usual way,
\begin{equation}\label{4.12}
  S=-\sum_{\bm{m}} P_s(\bm{m}) \ln P_s(\bm{m}) =
  \frac{N}{X} D_0^s + \ln Z \, ,
\end{equation}
which is an extensive quantity. The specific entropy per site is
\begin{eqnarray}
  \sigma & \equiv & \frac{S}{N}  =  \frac{1}{X} D_0^s +\ln\zeta \nonumber \\
   &  = &
  \frac{(1+4e^{-1/X})^{1/2}-1}{2X(1+4e^{-1/X})^{1/2}}+
  \ln \frac{1+(1+4e^{-1/X})^{1/2}}{2} \, . \nonumber \\
   \label{4.13} & &
\end{eqnarray}
It is possible to define a  function analogous to Helmholtz's specific
free energy,
\begin{equation}\label{4.14}
  \Phi=-X\ln \zeta= D_0^s -X\sigma \, ,
\end{equation}
so that Eq.~(\ref{4.9}) can be expressed
\begin{equation}\label{4.15}
  D_0^s=\frac{d(\Phi/X)}{d(1/X)} \, .
\end{equation}

The description of the steady state of the TIM and that of the tapped 1SFM
presented in Ref.\ \onlinecite{BPyS00} are closely related. In both
models, the metastable states are the same, namely those characterized by
having all the holes isolated, i.\ e.\/, surrounded by two particles.
Moreover, in the weak tapping limit, the steady state probability
distribution has the canonical form in both models. This implies that
their ``thermodynamical'' properties are the same, when expressed in terms
of the compactivity. Probably, this equivalence does not hold for stronger
tapping, for which  the simple description for the steady state developed
here seems to need some refinements,  as it follows from the numerical
experiments reported in Refs.~\onlinecite{BFyS02,Le02}. Finally, let us
note that the number of holes must be even in the TIM, while the 1SFM is
free from this restriction. Of course,  this difference becomes irrelevant
in the thermodynamic limit, in which the density of holes is a continuous
variable.

\subsection*{Spatial Correlations}

In the steady state, the only spatial correlations present in the model are due
to the impossibility of having two nearest neighbour holes. This property can be
used to simplify the calculation of the correlation functions in the steady
state.

We introduce two ``entities'', $\alpha$ and $\beta$, being $\alpha$ the
set formed by a hole with the particle that is at its right nearest
neighbour site, and $\beta$ a single particle. Thus, any metastable
configuration is obtained as an, unrestricted, arbitrary permutation of
the entities $\alpha$ and $\beta$. Here we take a very large system, so
that we do not need to consider periodic boundary conditions. We will
denote by $N_\alpha$ and $N_\beta$ the number of entities $\alpha$ and
$\beta$ in a given configuration, respectively. Therefore,
\begin{equation}\label{4.16}
  2N_\alpha+N_\beta=N \, .
\end{equation}
It is obvious that $N_\alpha$ equals the number of holes $k$ in the
configuration, so that the density of entities $\alpha$ and $\beta$ are
related to the density of holes $D_0$,
\begin{subequations}\label{4.17}
\begin{equation}\label{4.17a}
  x_\alpha\equiv\frac{\langle N_\alpha \rangle}{N}=D_0 \, ,
\end{equation}
\begin{equation}\label{4.17b}
  x_\beta\equiv\frac{\langle N_\beta \rangle}{N}=1-2x_\alpha=1-2D_0 \, ,
\end{equation}
\end{subequations}
where the averages are done over the considered ensemble of systems.

Since the positions of entities $\alpha$ and $\beta$ are independent in
the steady state, it is very easy to compute stationary correlations. For
instance,  the probability of finding two holes separated by $r$ particles
is given by
\begin{equation}\label{4.18}
  F_r^s=\langle m_k (1-m_{k+1})\cdots(1-m_{k+r})m_{k+r+1}\rangle_s\equiv
  x^s_{\alpha\beta^{r-1}\alpha} \, .
\end{equation}
Here, $x^s_{\alpha\beta^{r-1}\alpha}$ is the steady density of clusters
composed by two entities $\alpha$ separated by $r-1$ entities $\beta$. The
number of entities $\beta$ is $r-1$ because the particle in site $k+1$
together with the hole in site $k$ constitute the first entity $\alpha$.
We will consider that $r\geq 1$, since the holes are isolated in the
metastable states. It is
\begin{equation}\label{4.19}
    x_{\alpha\beta^{r-1}\alpha}=
    \frac{\langle
    N_{\alpha\beta^{r-1}\alpha}\rangle}{N}
    \, ,
\end{equation}
where $N_{\alpha\beta^{r-1}\alpha}$ is the number of clusters of the kind
indicated above. By definition,
\begin{equation}\label{4.20}
\langle
N_{\alpha\beta^{r-1}\alpha}\rangle=
\langle N_\alpha \rangle P(\beta^{r-1}\alpha | \alpha)
\, ,
\end{equation}
where $P(\beta^{r-1}\alpha | \alpha)$ is the conditional probability of finding
a cluster composed of $r-1$ consecutive entities $\beta$ and one entity $\alpha$
to the right of one entity $\alpha$. As the entities $\alpha$ and $\beta$ are
independent in the steady state, the stationary value of this conditional
probability is
\begin{equation}\label{4.21}
 P_s(\beta^{r-1}\alpha | \alpha) =  P_s(\beta^{r-1}\alpha)=
 \left[ P^s_\beta \right]^{r-1} P^s_\alpha \,  ,
\end{equation}
where $P^s_\alpha$ and $P^s_\beta$ are the probabilities of finding an $\alpha$
and a $\beta$ entity in the steady state, respectively. Obviously,
\begin{subequations}\label{4.22}
\begin{equation}\label{4.22a}
  P^s_\alpha=\frac{\langle N_\alpha  \rangle^s}
  {\langle N_\alpha+N_\beta \rangle^s} =\frac{D_0^s}{1-D_0^s} \, ,
\end{equation}
\begin{equation}\label{4.22b}
  P^s_\beta=\frac{\langle N_\beta  \rangle^s}
  {\langle N_\alpha+N_\beta \rangle^s} =\frac{1-2D_0^s}{1-D_0^s} \, ,
\end{equation}
\end{subequations}
where we have made use of Eq.~(\ref{4.17}). Therefore, putting together
Eqs.~(\ref{4.18})-(\ref{4.22}), we get
\begin{equation}\label{4.23}
  F_r^s=\frac{\left( D_0^s \right)^2}{1-2 D_0^s}
  \left( \frac{1-2D_0^s}{1-D_0^s} \right)^r \, \quad (r\geq 1) \, .
\end{equation}
Since the moment $F_r^s$ equals the probability of finding two holes separated
by $r$ particles, it is clear that
\begin{equation}\label{4.24}
  F_0^s=0 \, ,
\end{equation}
reflecting that two holes must always be separated by at least one
particle. The moments $F_r$ obey the following ``sum rule'',
\begin{equation}\label{4.25}
  \sum_{r=0}^\infty F_r =D_0 \, ,
\end{equation}
expressing that the sum of the probabilities of finding two holes
separated by an arbitrary number of particles equals the probability of
finding one hole, i.\ e.\/,
\begin{eqnarray}
  D_0 & = & \langle m_k \rangle=
  \underbrace{\langle m_k m_{k+1}\rangle}_{\textstyle F_0}+
  \langle m_k (1-m_{k+1})\rangle \nonumber \\
   & = & F_0+F_1
  +\langle m_k (1-m_{k+1}) (1-m_{k+2})\rangle \nonumber \\
  & = & \cdots =\sum_{r=0}^\infty F_r \, .
\end{eqnarray}

The calculation of other spatial correlations in the steady state is
straightforward, by following a line of reasoning similar to the one used
to find $F_r^s$. For instance, $F_r^s$ also provides the probability of
finding a cluster composed by two entities $\alpha$ and $r-1$ entities
$\beta$, no matter the way they are ordered, because of the independence
of the entities $\alpha$ and $\beta$ in the steady state.

\section{Final remarks}
\label{s5}

In this paper we have analyzed a one-dimensional Ising model with nearest
neighbour interactions formulated in a way appropriated for  the study of
compaction in granular media. An equivalent particle-hole description has
been introduced, in which the holes are associated to the domain walls of
the original Ising system. The free relaxation of the system is modelled
by a $T=0$ dynamics \cite{LyD01}, which only allows those spin flips
decreasing the energy of the system. Any configuration with all the holes
(domain walls) being isolated is metastable, i.\ e, it does not evolve
with this $T=0$ dynamics. The tapping process is described as composed of
two steps: (a) vibration, i.\ e.\/, starting from a metastable
configuration, each spin of the system is flipped with probability $p$,
and (b)  the system freely relaxes with the $T=0$ dynamics until it
reaches a, in general, different metastable configuration. The parameter
characterizing the tapping process is the ``vibration intensity'' $p$.

In the particle-hole description, the dynamics at $T=0$ is analytically
solvable, by writing a closed hierarchy of equations for the probability
distribution functions $D_r$ of finding $r+1$ consecutive holes in the
system \cite{PyB01a}. In the long time limit, the system gets  stuck in a
state where $D_r=0$ for all $r>1$, i.\ e.\/, all the holes are isolated,
as indicated above.

Tapping is a rather complex process, since each tap is composed of two
neatly different processes: vibration and free relaxation. In order to get
a physical insight into the mechanisms responsible for the behaviour of
the system under tapping, the derivation of the {\it effective} transition
probabilities for the Markov process connecting the metastable states
reached by the system in two consecutive taps is needed. In general, this
is a formidable task, but in the limit of a gently tapped system these
transition rates between metastable states can be computed up to the
second order in $p$.

In the first order, the only possible transitions are diffusion and
destruction of a hole pair. Then, compaction takes place, since there are
no processes decreasing the density of particles in the system to the
lowest order. To describe the steady state, second-order processes must be
taken into account, so as to have transitions that increase the number of
holes. Interestingly, these transitions are just the inverse of those
decreasing the number of holes to the lowest order. Therefore, the  Markov
process is irreducible, i.\ e.\/, all the metastable configurations are
connected through a chain of transitions with non-zero probability. As a
consequence, there is an unique, well-defined, steady probability
distribution for each value of $p$. Besides, the effective transition
rates verify detailed balance. This property has been used to derive the
steady distribution analytically, finding that it has the canonical form.
Thus, a relationship between Edwards' compactivity and the vibration
intensity $p$ is obtained, in the limit of weak tapping.

The system analyzed in this paper, as formulated for modelling tapped
granular media, is closely related to the one-dimensional facilitated
Ising model \cite{BPyS99,BPyS00,PByS00}. In the respective particle-hole
pictures, the metastable states are the same, those having all the holes
isolated. Although, in the limit of weak tapping, the corresponding
effective dynamics connecting metastable configurations are not
equivalent, the steady state is described in both cases by the canonical
distribution. The role of the energy is played by the number of holes and
that of the temperature by Edwards' compactivity, which is related to the
vibration intensity by an Arrhenius-like expression.  In this way, we find
a sort of ``minimal'' model for Edwards' description of the steady state
of externally perturbed granular media: a one-dimensional system of
particles and holes, with the metastable states characterized by having
all the holes isolated, and a canonical probability distribution function.
Nevertheless, in order to have an  actually complete description of the
steady state, a relationship between the parameters characterizing the
statics and the dynamics of the system, i.\ e.\/, between the compactivity
and the vibration intensity, is needed. Then, it is also necessary to
derive the effective dynamics from the underlying original models (TIM,
1SFM, etc.\/), when trying to understand the steady state behaviour.

The effective dynamics approach between metastable states has been shown
to be a powerful tool, in order to study the steady state of models for
granular systems submitted to tapping processes. It allows to identify the
physical mechanisms responsible for the increase of the density, and also
for the existence of a steady state characterized by a density being a
monotonic decreasing function of the vibration intensity. In simple
models, the calculations can be thoroughly done in the limit of gentle
tapping, deriving analytically the steady state distribution. The results
so obtained are consistent with recent extensive numerical tests of
Edwards hypothesis in simple systems
\cite{BKLyS00,BKLyS01,LyD01,BFyS02,Le02}, although the systematic
deviations found for stronger tapping \cite{BFyS02,Le02} cannot be
accounted for within the second-order theory developed in this paper and
in Ref.~\onlinecite{BPyS00}. This would need an extension of the effective
dynamics approach to the whole range of vibration intensities, which is
certainly not an easy task. Nevertheless, it is hard to believe that the
simple structure of the transition probabilities of the effective dynamics
should remain unaltered for stronger tapping, i.\ e.\/, deviations from
the simple canonical distribution found here are to be expected.

\begin{acknowledgments}
We acknowledge support from the Ministerio de Ciencia y
Tecnolog\'{\i}a (Spain) through Grant No. BFM2002-00303 (partially
financed by FEDER funds) .
\end{acknowledgments}

\bibliography{references,myarticles}

\end{document}